\documentstyle[preprint,aps]{revtex}
\input epsf.tex
\begin{document}
\title{Finite Nuclei in a Relativistic Mean-Field Model \\
with Derivative Couplings}\author{M. Chiapparini}
\address{Departamento de F\'{\i}sica Nuclear e Altas
Energias \\
Centro Brasileiro de Pesquisas F\'{\i}sicas \\
Rua Xavier Sigaud 150, 22290-180 Rio de Janeiro RJ, Brazil}
\author{A. Delfino, M. Malheiro\thanks{Present address: Department of
Physics and Astronomy, University of Maryland, College Park, MD 20742,
USA}}
\address{Instituto de F\'{\i}sica - Universidade Federal
Fluminense \\
Outeiro de S\~ao Jo\~ao Batista s/n, \,24020-004 Centro, Niter\'oi \\
Rio de Janeiro, Brazil}
\author{A. Gattone}
\address{Departamento de F\'{\i}sica. Comisi\'on Nacional de
Energ\'{\i}a At\'omica, \\
Av. del Libertador 8250, 1429 Buenos Aires, Argentina}
\date{\today}
\maketitle
\begin{abstract}
We study finite nuclei, at the mean-field level, using the
Zimanyi-Moskowski model and one of its variations (the ZM3 model).  We
calculate energy levels and ground-state properties in nuclei where
the mean-field approach is reliable.  The role played by
the spin-orbit potential in sorting out mean-field model descriptions
is emphasized.
\end{abstract}


\section{Introduction}

At the mean-field level the linear $\sigma-\omega$ (or Walecka)
model~\cite{walecka} satisfactorily explains many properties of
nuclear matter and finite nuclei with two free parameters.  The
resulting nuclear-matter compression modulus at saturation density
exceeds, however, the experimental bound~\cite{brown-osnes,co}.  A way
out of this difficulty is to introduce non-linear scalar
self-couplings~\cite{boguta,waldhauser}.  The non-linear model that
obtains has been shown to reproduce well ground-state nuclear
properties (though with some instabilities at high densities and for
low values of the compressibility, $\kappa\lesssim 200$
MeV~\cite{waldhauser}).  This model is renormalizable and has four
free parameters to fit.

One alternative approach which renders a satisfactory compression
modulus without increasing the number of free parameters is that
advanced by Zimanyi and Moszkowski~\cite{zm} and by Heide and
Rudaz~\cite{heide}.  This model employs the same degrees of freedom
and the same number of independent couplings that are present in the
Walecka model.  The difference lies in the coupling among the fields.
In the work of Zimanyi and Moszkowski (ZM), for instance, the authors
use a non-renormalizable derivative coupling between the scalar-meson
and the baryon fields which they later adjust to reproduce the
experimental conditions at saturation.  Their results for the
compression modulus, $\kappa=224$ MeV, and effective mass, $M^*=797$
MeV, compare very well with Skyrme-type calculations~\cite{zimanyi}.

Since it is desirable to have models that, at the mean-field level,
provide reasonable nuclear matter results, the original ZM model and
its two variations (to which we shall refer hereafter as ZM2 and ZM3)
described in the appendix of the original paper of Zimanyi and
Moszkowski~\cite{zm}, have deserved some recent attention.
Particularly, Koepf, Sharma and Ring~\cite{koepf} (KSR) have performed
nuclear-matter and finite-nuclei calculations and compared the ZM
model against i) the linear $\sigma -\omega$ model and, ii) models
with non-linear scalar couplings of phenomenological origin.  One of
their findings was that the ZM model, when used to calculate the
energy spectrum of $^{208}$Pb, gives a spin-orbit splitting that
compares poorly with the data.  It is well known that the splitting of
the energy levels due to the spin-orbit interaction is very sensitive
to the difference $\lambda{=}V-S$ between the vector, $V$, and the
scalar, $S$, potentials.  For the ZM model one gets $\lambda{=}223$
MeV, which is short of the 785 MeV of the successful (in this regard)
Walecka model.  A simplified summary of KSR's results also indicates
that:  1) The (non)linear $\sigma -\omega$ model predicts a
(good)reasonable spin-orbit splitting, a (soft)stiff equation of state
and a (good)small effective nucleon mass; 2) The ZM model predicts a
soft equation of state, a reasonable effective nucleon mass and a poor
spin-orbit splitting for finite nuclei.  To this list we may
add~\cite{delfino,delfino2} that the relativistic content of the
various models, as given by the ratio of scalar to baryon
densities, differs also being the linear and non-linear
$\sigma-\omega$ models more relativistic than the usual ZM.

At this point it is clear that fixing the spin-orbit problem in the ZM
model would make it similar in results to the non-linear
$\sigma-\omega$ model but with the additional bonus of having to fit
two parameters instead of four.  The modified ZM models, ZM2 and ZM3,
aim at this.  All three models come about in the following way.  In
the standard ZM model the non-linear effective scalar factor
$1/m^*{=}1+g_\sigma \sigma/M$, whose meaning will become clear below,
multiplies the nucleon derivative and the nucleon-vector coupling
terms.  When its two free parameters are fitted to the nuclear-matter
baryon density, $\rho_0{=}0.148$ fm$^{-3}$, and energy density per
nucleon at saturation, $E_b{=}-15.75$ MeV, one obtains the results
shown in the first row of Table~\ref{table1}.  If we now let $1/m^*$
to act upon the nucleon derivative and all terms involving the vector
field we end up with the ZM2 version of the ZM model, whose
predictions are shown in the second row of the table.  The
compressibility is slightly lower and the difference between the
vector and scalar potentials has increased to 278 MeV~\cite{delfino2}.
Finally, the ZM3 version of the model is obtained by allowing $1/m^*$
to act just on the nucleon derivative term.  With this change one
obtains the results displayed in the third row of the table; the
compressibility diminishes again and $\lambda$ reaches 471
MeV~\cite{delfino2}.  It is worth calling the attention to the fact
that the particular non-linear couplings used in these models change
the traditional conception whereby to a lower value of $m^*=M^*/M$
should correspond a higher compressibility $\kappa$.  Particularly,
the ZM3 field equations couple the non-linear scalar field to the
vector field and this new source is responsible for the $m^*{=}0.72$
and $\kappa$=156 MeV, of the table.  In the phenomenological
non-linear scalar-coupling model of Feldmeir and
Lindner~\cite{feldmeier} (hiperbolic ansatz), for example, the price
to pay for the reduced effective mass, $m^*=0.71$, is an enhanced
compressibility, $\kappa=410$ MeV.

If we assume that the difference $\lambda$ between the scalar and the
vector potentials is a good indicator of the strength of the
spin-orbit splitting in finite nuclei, it is clear from
Table~\ref{table1} that ZM3 offers the closest approach of the three
to the non-linear $\sigma -\omega$ model.  In this situation it is
relevant to study the model predictions for a number of properties of
finite nuclei for which there exist experimental data.  In this work
we apply the ZM3 model to the study of the energy spectrum and
ground-state properties of $^{16}$O, $^{40}$Ca, $^{48}$Ca,
$^{90}$Zr and $^{208}$Pb.  We calculate energy levels, charge {\em
r.m.s} radius and nucleon density distributions and compare the
results obtained with the predictions of the standard ZM model and
with the experimental data in order to extract conclusions.

In the next section we introduce the ZM and ZM3 models and give the
necessary detail to understand the origin of the results we obtain.
These are presented in the last section where we also draw some
conclusions.

\section{The Model}

The ZM and ZM3 models of interest for this work can be derived
from the Lagrangean
\begin{eqnarray}
{\cal L} &=& \sum_{a=1}^A \bar{\psi}_a \left\{ \gamma_{\mu}
\left[\left(i\partial^{\mu}-e\:\frac{(1+\tau_3)}{2}
A^{\mu}\right)-g^*_{\omega}
\omega^{\mu}-\frac{g^*_{\rho}}{2}\tau_3 \rho^{\mu}\right]-
(M-g^*_\sigma\:\sigma)\right\}\psi_a
\label{L3}\\
&&+\frac{1}{2}\left(\partial_\mu\sigma\partial^\mu\sigma -
m^2_\sigma\sigma^2\right)
-\frac{1}{4}\left(\partial_\mu\omega_\nu-\partial_\nu\omega_\mu
\right)^2+
\frac{1}{2}m^2_\omega\omega_\mu\omega^\mu \nonumber\\
&&-\frac{1}{4}\left(\partial_\mu\rho_\nu-
\partial_\nu\rho_\mu\right)^2+
\frac{1}{2}m^2_\rho \rho_\mu \rho^\mu -\frac{1}{4}\left(\partial_\mu
A_\nu-
\partial_\nu A_\mu\right)^2\;, \nonumber
\end{eqnarray}
where the effective coupling constants are given in each model
according to
\begin{center}
\begin{tabular}{ccc}\hline\hline
{\rm Model}& $g^*_\sigma$  & $g^*_{\omega,\rho}$          \\ \hline
  {\rm ZM} & $m^*g_\sigma$ & $g_{\omega,\rho}$            \\
  {\rm ZM3}& $m^*g_\sigma$ & $m^*g_{\omega,\rho}$ \\ \hline\hline
\end{tabular}
\end{center}
where we define
\[
m^*=\left(1+ \frac{g_\sigma \sigma }{M}\right)^{-1}.
\]

In equation~(\ref{L3}) the nucleon field, $\psi$, couples to the
scalar-isoscalar meson field, $\sigma$, and the vector-isoscalar
meson
field, $\omega_\mu$.  A third isovector meson field, $\rho_\mu$,
(neutral component) is included to account for the asymmetry between
protons and neutrons.  The $\rho_\mu$ and the electromagnetic field
$A_\mu$ both couple to the baryon field.

The Euler-Lagrange equations of motion using the Lagrangean
(\ref{L3}) give the following equations for the fields,
\begin{eqnarray}
\left\{ \gamma_{\mu}
\left[\left(i\partial^{\mu}-e\:\frac{(1+\tau_3)}{2}
A^{\mu}\right)-g^*_{\omega}
\omega^{\mu}-\frac{g^*_{\rho}}{2}\tau_3 \rho^{\mu}\right]-
(M-g^*_\sigma\:\sigma)\right\}\psi_a &=&0 \;,
\end{eqnarray}
\begin{eqnarray}
\partial_\mu \omega^{\mu\nu} + m_\omega^2 \omega^\nu &=&
g^*_\omega \sum_{a=1}^A \bar{\psi}_a \gamma^\nu \psi_a \;,  \\
\partial_\mu \rho^{\mu\nu} + m_\rho^2\rho^\nu  &=&
\frac{g^*_\rho}{2} \sum_{a=1}^A \bar{\psi}_a
\gamma^\nu\tau_3 \psi_a \;, \\
\partial_\mu A^{\mu\nu} &=&
\frac{e}{2}\sum_{a=1}^A \bar{\psi}_a \gamma^\nu(1+\tau_3) \psi_a \;,
\\
\partial_\mu \partial^\mu \sigma + m_\sigma^2 \sigma &=&
{m^*}^2 g_\sigma\sum_{a=1}^A \bar{\psi}_a\psi_a
-\sum_{a=1}^A \left(\frac{\partial g^*_\omega}{\partial \sigma}
\bar{\psi}_a\gamma_\mu\psi_a\omega^\mu
+\frac{1}{2}\frac{\partial g^*_\rho}{\partial \sigma}
\bar{\psi}_a\gamma_\mu\tau_3\psi_a\rho^\mu \right) \;.
\end{eqnarray}

In the mean-field approximation all baryon currents are replaced by
their ground-state expectation values.  In a system with spherical
symmetry the mean value of the spatial components of the vector-meson
fields vanish, resulting in the following mean-field equations,
\begin{eqnarray}
\left\{ \gamma_{\mu}
i\partial^{\mu}-\gamma_0\left[e\:\frac{(1+\tau_3)}{2}A^0
-g^*_{\omega}\omega^0 -\frac{g^*_{\rho}}{2}\tau_3 \rho^0\right]-
(M-g^*_\sigma\:\sigma)\right\}\psi_a &=&0 \;,\label{dirac3}
\end{eqnarray}
\begin{eqnarray}
-\nabla^2 \omega^0 + m_\omega^2
\omega^0 &=& g^*_\omega \sum_{a=1}^A \psi^\dagger_a\psi_a \equiv
g^*_\omega \rho_b = g^*_\omega (\rho_p+\rho_n) \;,\label{omega3} \\
-\nabla^2 \rho^0 + m_\rho^2\rho^0 &=& \frac{g^*_\rho}{2} \sum_{a=1}^A
\psi^\dagger_a \tau_3 \psi_a \equiv g^*_\rho \rho_3 = g^*_\omega
(\rho_p-\rho_n) \;,\label{rho3}\\
-\nabla^2 A^0 &=&
\frac{e}{2}\sum_{a=1}^A \psi^\dagger_a(1+\tau_3) \psi_a \equiv e
\rho_p \label{coulomb} \;,\\ -\nabla^2 \sigma + m_\sigma^2 \sigma &=&
{m^*}^2 g_\sigma\sum_{a=1}^A \bar{\psi}_a\psi_a -\sum_{a=1}^A
\left(\frac{\partial g^*_\omega}{\partial \sigma}
\psi^\dagger_a\psi_a\omega^0 +\frac{1}{2}\frac{\partial
g^*_\rho}{\partial \sigma} \psi^\dagger_a\tau_3\psi_a\rho^0 \right)
\;, \label{sigma3}\\
&\equiv& {m^*}^2 g_\sigma \rho_s - \frac{\partial
g^*_\omega}{\partial \sigma} \rho_b \omega^0 - \frac{\partial
g^*_\rho}{\partial \sigma} \rho_3 \rho^0 \;.\nonumber
\end{eqnarray}

Equations~(\ref{dirac3})~to~(\ref{sigma3}) are a set of coupled
non-lineal differential equations which may be solved by iteration.
For a given set of initial meson potentials, the Dirac equation
(\ref{dirac3}) is solved.  Once the baryon wave functions are
determined the source terms for Eqs.~(\ref{omega3})-(\ref{sigma3})
are evaluated and new meson fields are obtained.  The procedure is
iterated until self-consistency is achieved.

The coupling constants $g_{\sigma}$ and $g_{\omega}$ are chosen to
reproduce the saturation baryon density in symmetric nuclear matter,
$\rho_0{=}0.148$ fm$^{-3}$, and the energy density per nucleon at
saturation, $E/A{=}-15.75$ MeV.  The third coupling constant,
$g_\rho$, is obtained by fitting the bulk symmetry energy
in nuclear matter given by the expression
\begin{equation}
a_4=\frac{g_\rho^2}{8m_\rho^2}\left(\frac{g^*_\rho}{g_\rho}\right)\rho_0 +
\frac{k_F^2}{6E^*_F}
\end{equation}
to $a_4{=}32.5$ MeV at $\rho{=}\rho_0$. We have used
\[ \rho_0=\frac{2}{3\pi^2}k_F^3 \]
and
\[ E_F^*=\sqrt{k_F^2+{m^*}^2M^2} \;.\]
The coupling constants and masses used in the calculations are given
in Table~{\ref{table2}}.

\section{Results}

Results of the present calculation for the single particle energy
levels in $^{16}$O, $^{40}$Ca, $^{48}$Ca, $^{90}$Zr, and $^{208}$Pb
are shown in Tables~\ref{table3} to~\ref{table5}.  Through the tables
a good overall agreement is obtained, i) between the two models and,
ii) with the experimental data.  Nonetheless, the ZM3 model fares
better than ZM in reproducing the observed spin-orbit splittings.
This is due to the difference in the way that the scalar and vector
mesonic fields couple to each other ---via non-linear effective
coupling constants--- in each model.  This difference is better
illustrated in figures~\ref{fig1} to~\ref{fig4} where the ``central
potential", $V_0$, and the ``reduced spin-orbit term", $V_{ls}$, are
depicted as functions of the radial distance, $r$, for oxygen and
lead.  We have defined the central potential to be
\begin{equation}
V_0=  V+S       \label{V0}
\end{equation}
with
\[ V = g_\omega^* \omega_0 \] and
\[ S = -g_\sigma^* \sigma. \] The reduced spin-orbit term, on the
other hand, is given by the expression,
\begin{equation}
V_{ls} =\frac{V' - S'}{2m_sr} \;, \label{vls}
\end{equation}
with $ m_s = M - 0.5*( V - S )$ and with the primes denoting partial
derivatives with respect to $r$.  Though the choice of mass in the
denominator of Eq.~(\ref{vls}) could be opened to some debate, the
figures are intended to compare strengths in both models and not to
draw absolute conclusions.

Figures~\ref{fig1} and~\ref{fig3} illustrate the origin of the
behaviour of both models.  ZM is shallower in the center and raises
more steeply on the surface than ZM3; this holds true both in oxygen
and in lead.  The resulting spin-orbit potential (figures~\ref{fig2}
and~\ref{fig4}) is, thus, deeper at the surface in ZM3 than in ZM.
To estimate the magnitude of this difference we calculated the
ratio expected for the spin-orbit splittings in ZM3 and ZM, using the
parameters from the fit to nuclear matter of Table~\ref{table1}.  The
result \[ \frac{V^{ZM3}_{ls}}{V^{ZM}_{ls}} \propto \frac{(V-S)^{ZM3}
m^*_{ZM}} {(V-S)^{ZM} m^*_{ZM3}} \approx 2.5 \] agrees nicely with the
ratios extracted from the levels in tables~\ref{table3}-\ref{table5}.

It is worth noticing that our calculated spectra for the ZM model,
differ from those presented by KSR in Ref.~\cite{koepf} for the same
model.  We have traced the origin of this discrepancy to the symmetry
energy $a_4$ used to fit the $\rho$-meson coupling constant.
Figure~\ref{fig5} illustrates the effect that a decreasing $a_4$ has
on the energy of a few selected single particle orbits in $^{208}$Pb.
The shift in energy due to the presence of the $\rho$-meson may be as
large as 5 MeV depending on the choice of $a_4$ (the arrow indicates
the value used in the calculations shown here, $a_4=32.5$ MeV).  The
spectrum of $^{208}$Pb calculated with a vanishing $g_{\rho}$ agrees
with the results of KSR.  Notice also that the splittings amongst
spin-orbit partners are insensitive to $g_{\rho}$ since changes in
$a_4$ shift the levels globally.

At this point one of our conclusions is that, as demonstrated
in~\cite{koepf}, the ZM model does not do well in reproducing the
energy splittings due to the spin-orbit interaction.  However, the
overall spectrum turns out to be satisfactory.  In this sense the
isovector meson plays in ZM the same important role, for asymmetric
nuclei, that has been shown to play in the non-linear Walecka
model~\cite{sharma}.

Table~\ref{table6} shows the results obtained for some of the static
ground-state properties of the same nuclei as above.  One feature to
point out is that, systematically, ZM3 predicts a {\em r.m.s.} value
for the charge radius that is larger than the one calculated with the
ZM model.  This is due to the slightly smaller binding of the protons
in this model that produces a longer tail in the charge distribution,
as shown in Figs.~\ref{fig6} and~\ref{fig7}.  The baryon density
extends also farther in ZM3 as can be appreciated from
Figs.~\ref{fig8} and~\ref{fig9}.  The baryon distribution at the
surface indicates that edge effects are more important in ZM3 than in
the ZM model, something that agrees with our discussion of the
spin-orbit splittings in the previous paragraphs.

Summarizing, we have applied the ZM3 model to the study of the energy
spectrum and ground-state properties of $^{16}$O, $^{40}$Ca,
$^{48}$Ca, $^{90}$Zr and $^{208}$Pb.  The interest in this model, and
the standard ZM, resides in their ability to describe nuclear matter
at saturation with two free parameters.  For finite nuclei, we have
shown that ZM3 gives a reasonable description of nuclear spectra and
improves upon ZM regarding the energy splitting of levels due to the
spin-orbit interaction.  The results of the calculations described
here were compared with those obtained using the standard ZM model and
with the experimental data.  It is our conclusion that models of this
kind, with derivative couplings involving the scalar and the vector
fields, offer a valid framework to pursue calculations where the
requirement of simultaneous reasonable values for the nuclear
compressibility and the spin-orbit splitting cannot be side-stepped.

\vspace{1cm}

{\bf Acknowledgements}

One of the authors (MC) would like to express his thanks to the
Conselho Nacional de Desenvolvimento Cient\'{\i}fico e Tecnol\'ogico
(CNPq) and to the Departamento de Fisica Nuclear e Altas Energias
(DNE-CBPF) Brazil, for their financial support.  AG is a fellow of the
the CONICET, Argentina.

\newpage

\begin{table}
\caption{Nuclear matter results in the three Zimanyi-Moskowszki
models.  The compressibility $\kappa$, the scalar potential $S$,
and the vector potential $V$, are in MeV. The effective mass is in
units of the bare mass.  }
\begin{tabular}{ccccc}
{\rm Model}& $\kappa$ &{\rm S}   &   {\rm V}   &   $m^*$   \\ \hline
  {\rm ZM} & 225. &   -141.   &   82.5     &   0.85    \\
  {\rm ZM2}& 198. &   -168.   &   110.7    &   0.82    \\
  {\rm ZM3}& 156. &   -267.   &   204.7    &   0.72    \\
\end{tabular}
\label{table1}
\end{table}

%\newpage
\begin{table}
\caption{Parameters used in the calculation.}
\begin{tabular}{cccccccc}
{\rm
Model}&$m_N$&$m_\sigma$&$m_\omega$&$m_\rho$&$g_\sigma^2$&$g_\omega^2$&$g_\rho^2$
\\ \hline
{\rm ZM}   &938.27& 520     & 783      & 770    & 55.540     & 44.207
    & 79.605    \\
{\rm ZM3}  &938.27& 520     & 783      & 770    & 135.30     & 211.21
    & 137.74    \\
\end{tabular}
\label{table2}
\end{table}

% \ \vspace*{20cm}

\begin{table}
\caption{Single-particle energy levels (in MeV) calculated in the
Zimanyi-Moszkowski model (ZM) and in the ZM3 version.  Where
available
the experimental energies are also quoted. Numbers between brackets
correspond to protons, others to neutrons.}
\begin{tabular}{cccrrcrrcrr}
{\rm Nucl.}   &{\rm Levels}& &\multicolumn{2}{c}{\rm ZM}& &
                             \multicolumn{2}{c}{\rm ZM3}& &
                             \multicolumn{2}{c}{\rm Exp.}   \\\hline
$^{16}${\rm O}  &$1s_{1/2}$& &35.4&(31.2)& &36.2&(32.1)&
&47.0&(40$\pm$8)\\
                &$1p_{3/2}$& &19.6&(15.6)& &19.4&(15.6)& &21.8&(18.4)
\\
                &$1p_{1/2}$& &18.2&(14.2)& &16.5&(12.7)& &15.7&(12.1)
\\\hline
$^{40}${\rm Ca} &$1s_{1/2}$& &43.3&(35.1)& &46.7&(38.7)& &
&(50$\pm$10)\\
                &$1p_{3/2}$& &32.4&(24.5)& &33.6&(25.9)& &
&(34$\pm$6)\\
                &$1p_{1/2}$& &31.6&(23.7)& &31.9&(24.2)& &
&(34$\pm$6)\\
                &$1d_{5/2}$& &20.2&(12.6)& &20.4&(13.0)& &21.9&(15.5)
   \\
                &$1d_{3/2}$& &18.7&(11.2)& &17.3&(9.97)& &15.6&(8.3)
   \\
                &$2s_{1/2}$& &15.9&(8.33)& &16.9&(9.52)& &18.2&(10.9)
  \\\hline
$^{48}${\rm Ca} &$1s_{1/2}$& &41.8&(39.9)& &45.7&(43.2)& &
&(55$\pm$9)\\
                &$1p_{3/2}$& &31.3&(30.7)& &33.2&(32.1)& &
&(35$\pm$7)\\
                &$1p_{1/2}$& &30.8&(30.1)& &32.0&(30.7)& &
&(35$\pm$7)\\
                &$1d_{5/2}$& &19.8&(19.7)& &20.4&(20.0)& &16.0&(20.0)
   \\
                &$1d_{3/2}$& &18.7&(18.5)& &17.9&(17.4)& &12.4&(15.3)
   \\
                &$1f_{7/2}$& &7.87&      & &7.94&      & &9.9 &
   \\
                &$2s_{1/2}$& &16.7&(14.4)& &17.2&(15.4)& &12.4&(15.8)
   \\
\end{tabular}
\label{table3}
\end{table}

\begin{table}
\caption{Same as Table \protect\ref{table3}.}
\begin{tabular}{cccrrcrrcrr}
{\rm Nucl.}   &{\rm Levels}& &\multicolumn{2}{c}{\rm ZM}& &
                             \multicolumn{2}{c}{\rm ZM3}& &
                             \multicolumn{2}{c}{\rm Exp.}   \\\hline

$^{90}${\rm Zr} &$1s_{1/2}$& &45.9&(37.2)& &51.0&(41.6)& &
&(54$\pm$8)\\
                &$1p_{3/2}$& &38.4&(30.8)& &41.6&(33.5)& &
&(43$\pm$8)\\
                &$1p_{1/2}$& &38.1&(30.5)& &40.9&(32.7)& &
&(43$\pm$8)\\
                &$1d_{5/2}$& &29.7&(22.9)& &31.4&(24.3)& &
&(27$\pm$8)\\
                &$1d_{3/2}$& &29.1&(22.2)& &29.9&(22.6)& &
&(27$\pm$8)\\
                &$1f_{7/2}$& &20.3&(13.8)& &20.8&(14.3)& &    &
   \\
                &$1f_{5/2}$& &19.1&(12.6)& &18.3&(11.7)& &13.5&
   \\
                &$1g_{9/2}$& &10.2&      & &10.2&      & &12.0&
   \\
                &$2s_{1/2}$& &26.6&(18.2)& &28.1&(19.9)& &
&(27$\pm$8)\\
                &$2p_{3/2}$& &16.0&(7.64)& &16.9&(8.88)& &13.1&
   \\
                &$2p_{1/2}$& &15.6&(7.25)& &16.0&(8.01)& &12.6&
   \\
\end{tabular}
\label{table4}
\end{table}

\begin{table}
\caption{Same as Table \protect\ref{table3}.}
\begin{tabular}{cccrrcrrcrr}
{\rm Nucl.}   &{\rm Levels}& &\multicolumn{2}{c}{\rm ZM}& &
                              \multicolumn{2}{c}{\rm ZM3}& &
                              \multicolumn{2}{c}{\rm Exp.} \\\hline
$^{208}${\rm Pb}&$1s_{1/2}$& &45.1&(35.6)& &50.6&(40.0)& &    &
\\
                &$1p_{3/2}$& &41.0&(31.8)& &45.3&(35.2)& &    &
\\
                &$1p_{1/2}$& &40.7&(31.6)& &45.1&(34.9)& &    &
\\
                &$1d_{5/2}$& &35.8&(26.9)& &39.0&(29.4)& &    &
\\
                &$1d_{3/2}$& &35.6&(26.6)& &38.4&(28.7)& &    &
\\
                &$1f_{7/2}$& &29.9&(21.2)& &31.9&(22.7)& &    &
\\
                &$1f_{5/2}$& &29.4&(20.6)& &30.9&(21.5)& &    &
\\
                &$1g_{9/2}$& &23.2&(14.6)& &24.3&(15.4)& &    &
\\
                &$1g_{7/2}$& &22.4&(13.8)& &22.6&(13.6)& &
&(11.4)\\
                &$1h_{11/2}$ &&15.9&(7.41)& &16.2&(7.65)& &    &(9.4)
\\
                &$1h_{9/2}$& &14.8&      & &13.9&      & &10.8&
\\
                &$1I_{13/2}$& &8.18&      & &7.98 &     & &9.0 &
\\
                &$2s_{1/2}$& &33.2&(23.7)& &36.2&(26.1)& &    &
\\
                &$2p_{3/2}$& &25.8&(16.5)& &27.8&(18.1)& &    &
\\
                &$2p_{1/2}$& &25.6&(16.3)& &27.4&(17.6)& &    &
\\
                &$2d_{5/2}$& &18.1&(8.87)& &19.4&(9.92)& &    &(9.7)
\\
                &$2d_{3/2}$& &17.7&(8.50)& &18.5&(9.06) & &    &(8.4)
\\
                &$2f_{7/2}$& &10.2&      & &11.0&      & &9.7 &
\\
                &$2f_{5/2}$& &9.69&      & &9.85&      & &7.9 &
\\
                &$3s_{1/2}$& &16.3&(6.75)& &17.6&(7.81) & &    &(8.0)
\\
                &$3p_{3/2}$& &7.88&      & &8.83&      & &8.3 &
\\
                &$3p_{1/2}$& &7.69&      & &8.38&      & &7.4 &
\\
\end{tabular}
\label{table5}
\end{table}

\begin{table}

\caption{The binding energy per nucleon $\varepsilon$ (in MeV), the
mean square charge radius $\langle r^2 \rangle$ (in fm$^2$) and, the
spin-orbit splitting (for partners where data are available)
$\Delta E_{so}$ (MeV), are shown for the ZM and ZM3 models.  The
experimental values are shown for reference.}

\begin{tabular}{ccccc}
{\rm Nucleus}&g.s.p.&{\rm ZM}&{\rm ZM3}&{\rm Exp.}\\ \hline
$^{16}{\rm O}$&$\varepsilon$&-8.40&-7.50&-7.98\\
              &$\langle r^2 \rangle$  &2.64 & 2.78& 2.74\\
&$\Delta E_{so}$($1p_{3/2}$-$1p_{1/2}$)&1.4&2.8&6.1\\ \hline
$^{40}{\rm Ca}$&$\varepsilon$&-9.01&-8.43&-8.55\\
               &$\langle r^2 \rangle$   & 3.39& 3.51&3.48\\
&$\Delta E_{so}$($1d_{5/2}$-$1d_{3/2}$)&1.5&6.1&6.3\\ \hline
$^{48}{\rm Ca}$&$\varepsilon$&-8.78&-8.30&-8.67 \\
               &$\langle r^2 \rangle$   & 3.47& 3.57&3.47 \\
&$\Delta E_{so}$($1d_{5/2}$-$1d_{3/2}$)&1.1&2.5&3.6\\ \hline
$^{90}{\rm Zr}$&$\varepsilon$&-8.80&-8.46&-8.71 \\
               &$\langle r^2 \rangle$   & 4.25& 4.35&4.27\\ \hline
&$\Delta E_{so}$($2p_{3/2}$-$2p_{1/2}$)&0.4&0.9&0.5\\
$^{208}{\rm Pb}$&$\varepsilon$&-7.86&-7.66&-7.87\\
               &$\langle r^2 \rangle$   & 5.54& 5.66& 5.50 \\
&$\Delta E_{so}$($2f_{7/2}$-$2f_{5/2}$)&0.5&1.15&1.8\\
&$\Delta E_{so}$($3p_{3/2}$-$3p_{1/2}$)&0.19&0.45&0.9\\
\end{tabular}
\label{table6}
\end{table}

\newpage
\begin{figure}

\caption{The potential $V_0$ (defined in equation (\protect\ref{V0})
in the text) as a function of the radial distance. The calculation
was done in oxygen and the results correspond to the standard
Zimanyi-Moszkowski model (ZM)~\protect\cite{zm} and to the ZM3
model studied in this paper. }
\label{fig1}
\end{figure}
\begin{figure}
\caption{Same as figure~\protect\ref{fig1} but for the $^{208}$Pb
nucleus.}
\label{fig2}
\end{figure}

\begin{figure}
\caption{The spin-orbit potential $V_{ls}$ (defined in equation
(\protect\ref{vls})
in the text) as a function of the radial distance. The calculation
was done in oxygen and the results correspond to the standard
Zimanyi-Moszkowski model (ZM)~\protect\cite{zm} and to the ZM3
model studied in this paper.}
\label{fig3}
\end{figure}

\begin{figure}
\caption{Same as figure~\protect\ref{fig3} but for the $^{208}$Pb
nucleus.}
\label{fig4}
\end{figure}
\begin{figure}
\caption{Valence proton and neutron orbitals in $^{208}$Pb as a
function of the assymetry parameter $a_4$. The arrow indicates the
value used in the calculations performed in this work.}
\label{fig5}
\end{figure}
\begin{figure}
\caption{The charge density distribution in $^{16}$O
as a function of the radial distance.
The results correspond to the standard
Zimanyi-Moszkowski model (ZM)~\protect\cite{zm} and to the ZM3
model studied in this paper.}
\label{fig6}
\end{figure}
\begin{figure}
\caption{Same as figure~\protect\ref{fig6} but for the $^{208}$Pb
nucleus.}
\label{fig7}
\end{figure}
\begin{figure}
\caption{  The baryon density distribution in $^{16}$O
as a function of the radial distance.
The results correspond to the standard
Zimanyi-Moszkowski model (ZM)~\protect\cite{zm} and to the ZM3
model studied in this paper.}
\label{fig8}
\end{figure}
\begin{figure}
\caption{Same as figure~\protect\ref{fig8} but for the $^{208}$Pb
nucleus.}
\label{fig9}
\end{figure}

\newpage

\begin{centering}
\epsfysize=21.5truecm
\epsffile{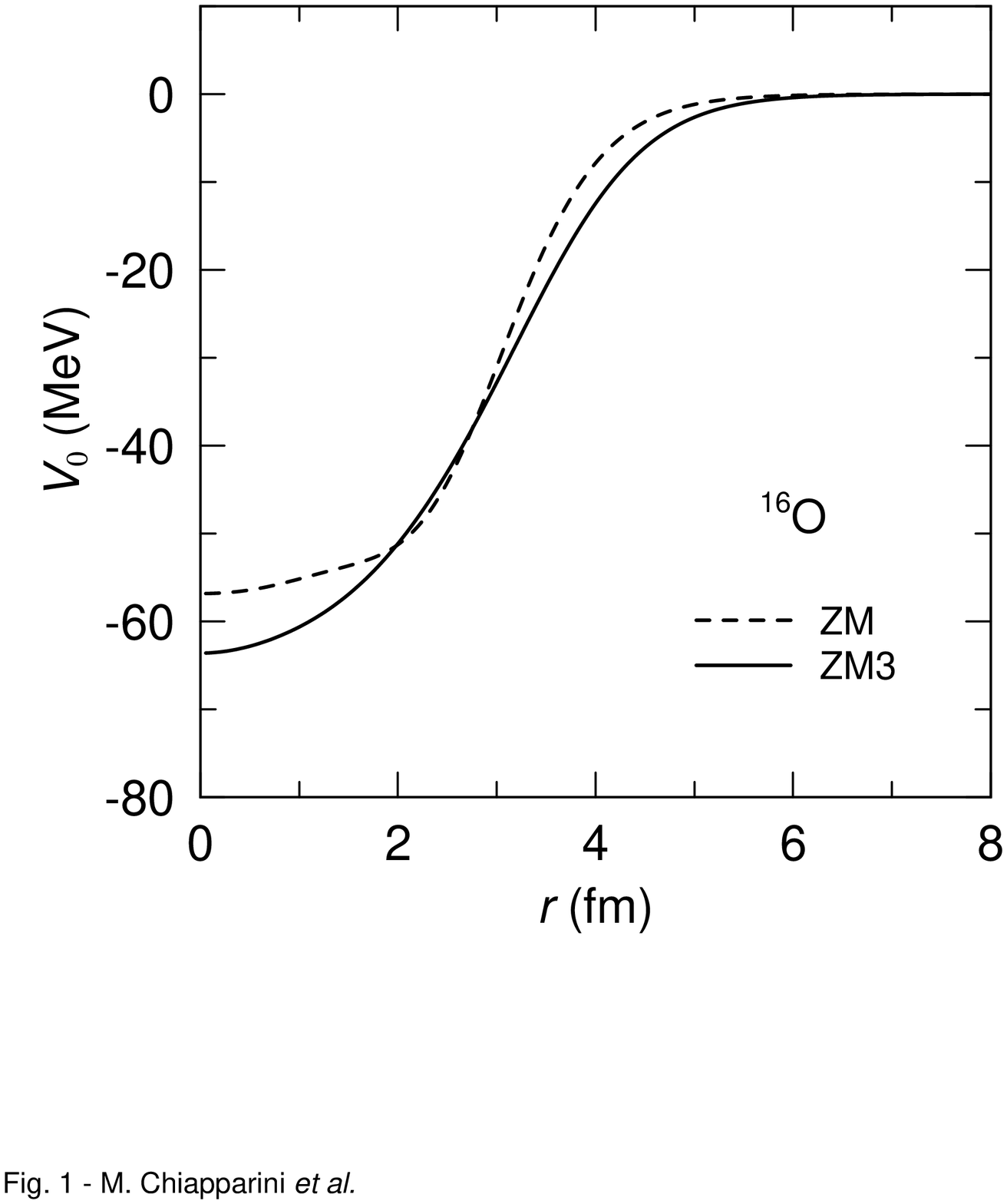}
\end{centering}

\newpage

\begin{centering}
\epsfysize=21.5truecm
\epsffile{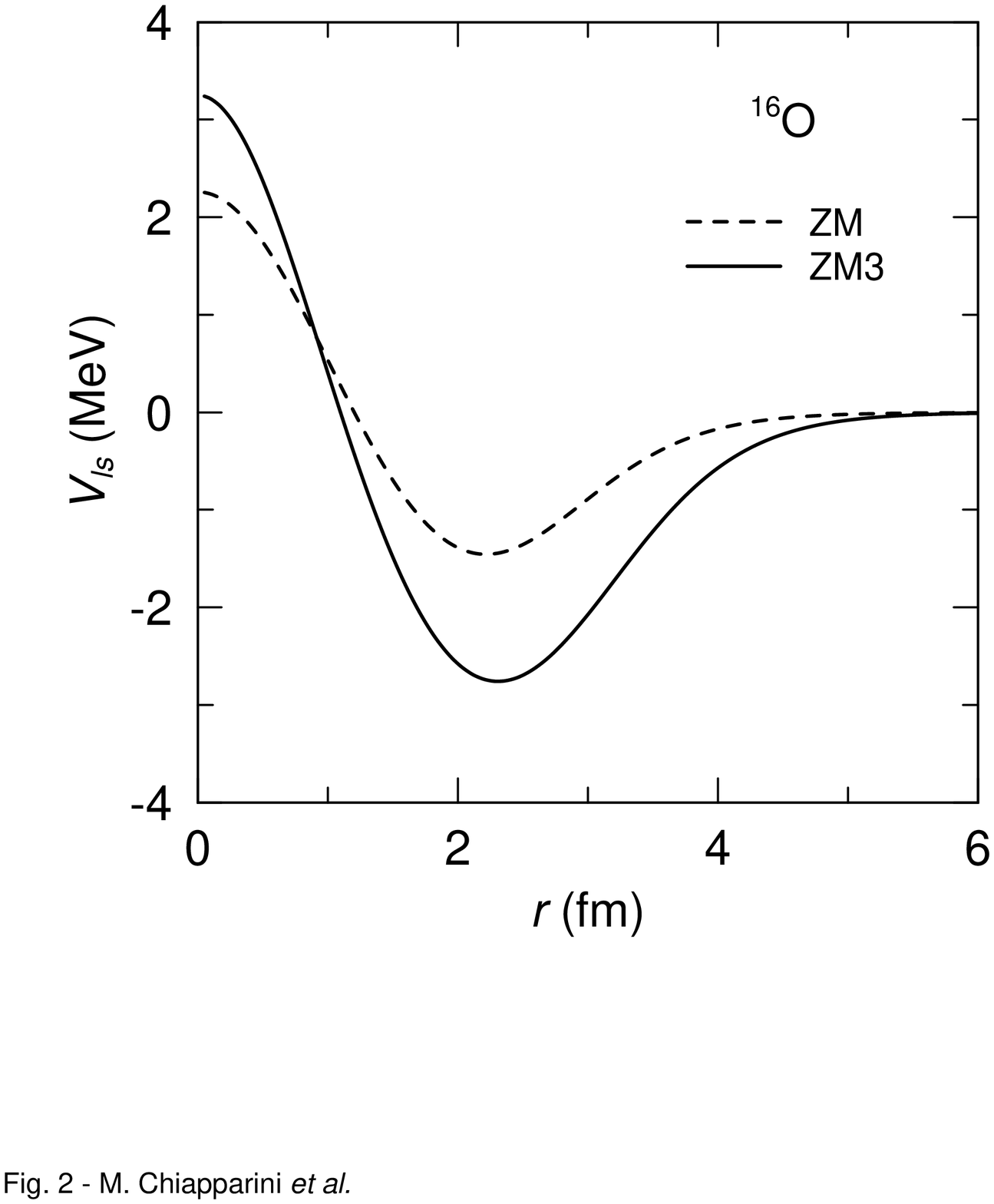}
\end{centering}

\newpage

\begin{centering}
\epsfysize=21.5truecm
\epsffile{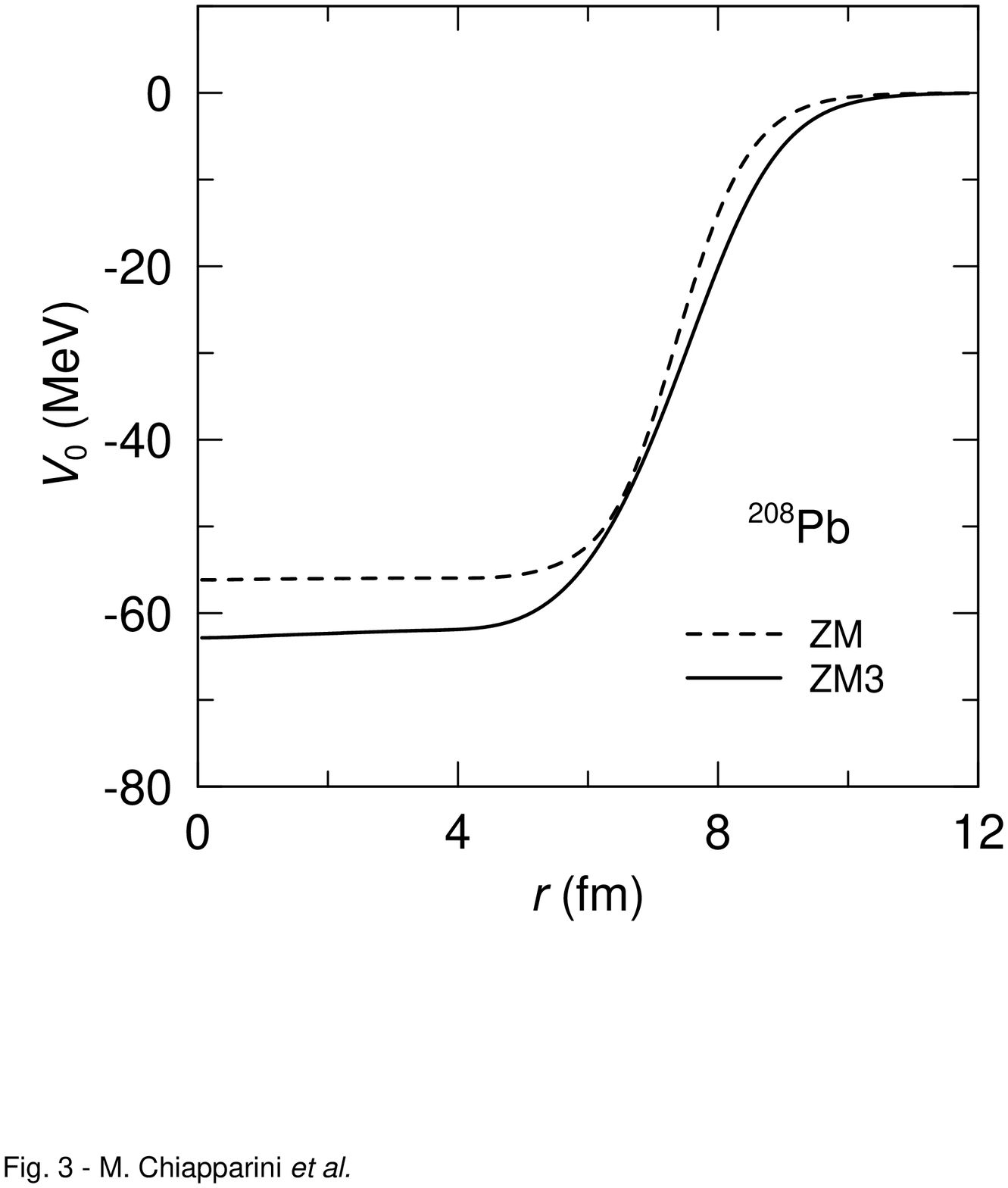}
\end{centering}

\newpage

\begin{centering}
\epsfysize=21.5truecm
\epsffile{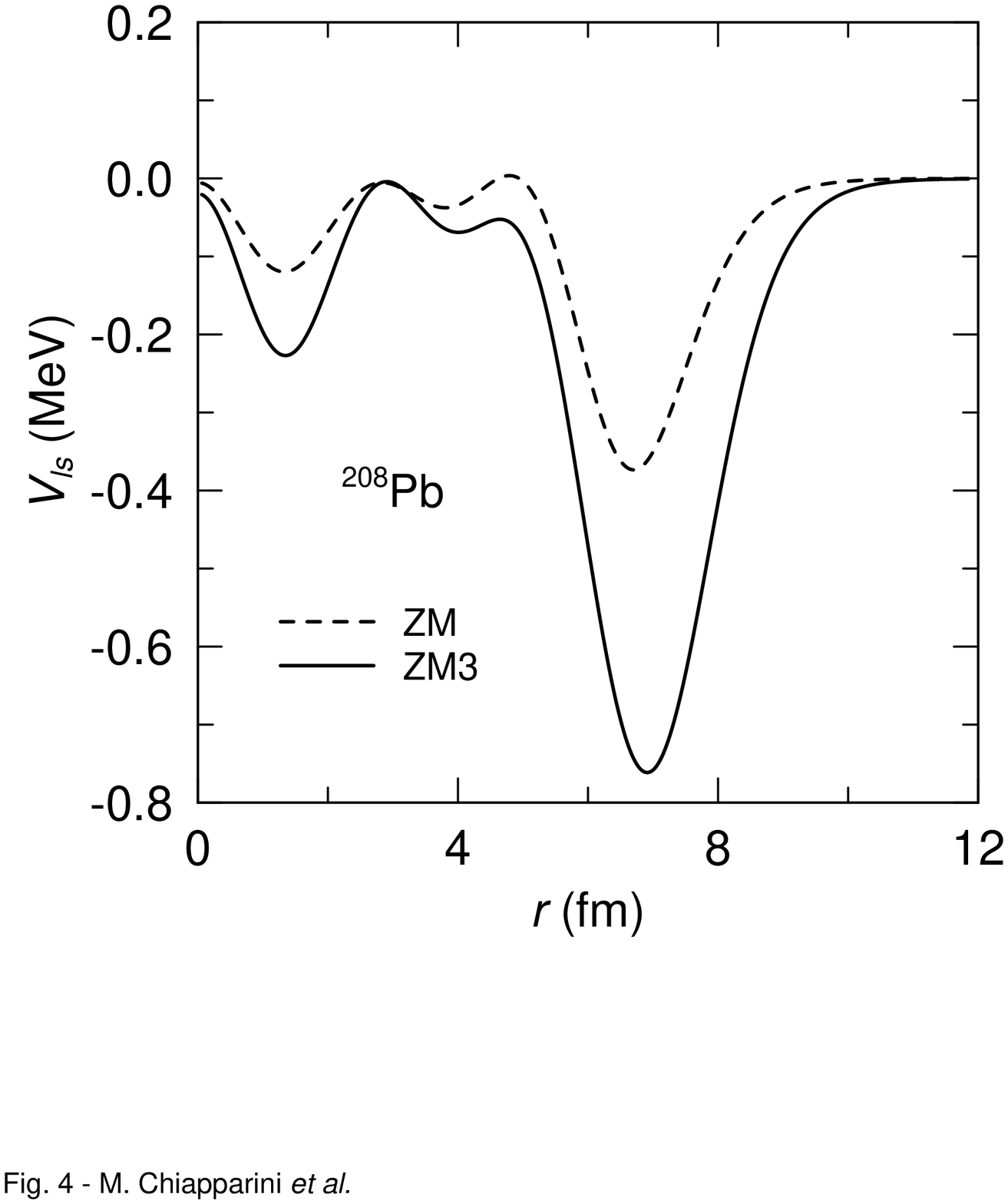}
\end{centering}

\newpage

\begin{centering}
\epsfysize=21.5truecm
\epsffile{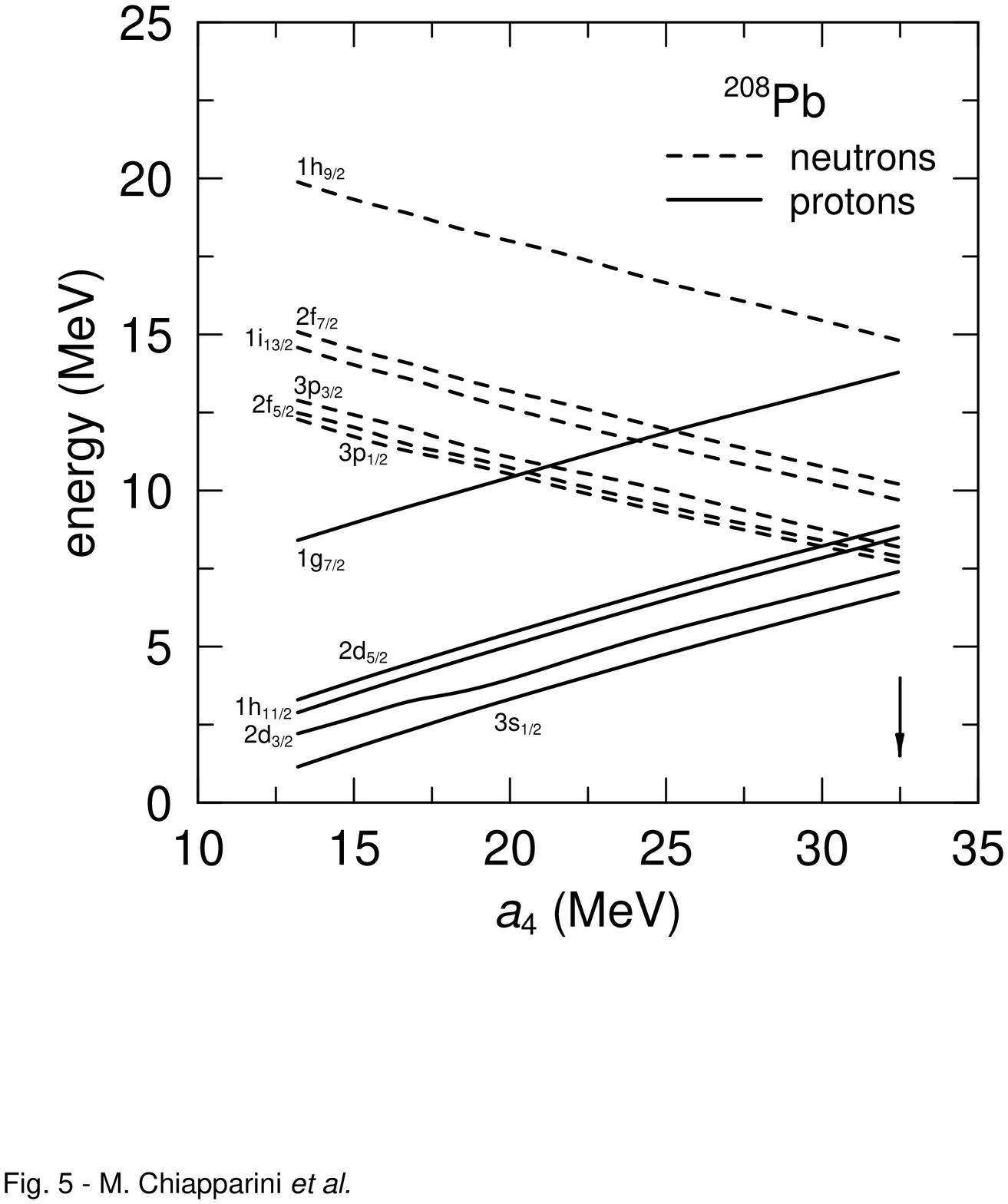}
\end{centering}

\newpage

\begin{centering}
\epsfysize=21.5truecm
\epsffile{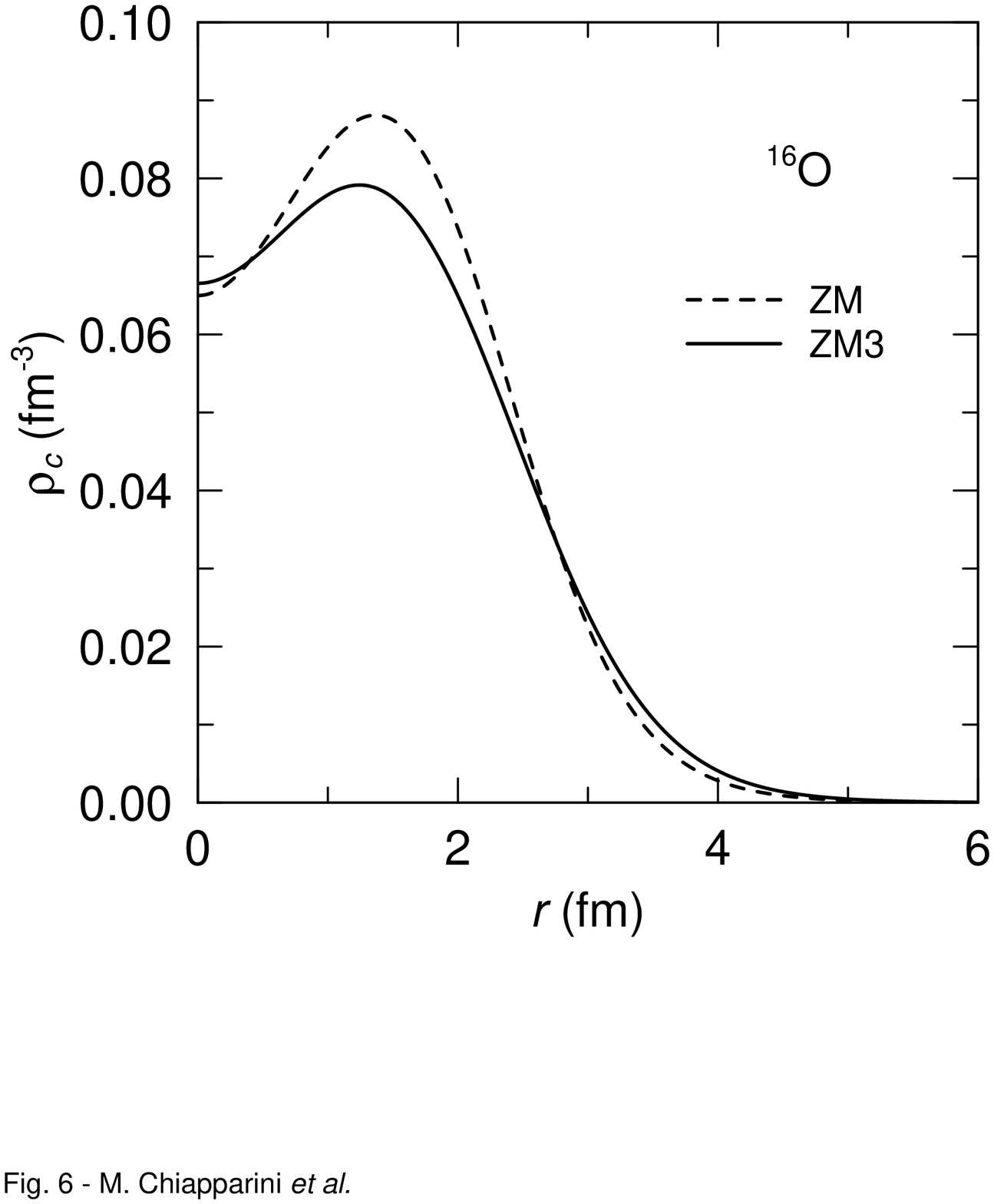}
\end{centering}

\newpage

\begin{centering}
\epsfysize=21.5truecm
\epsffile{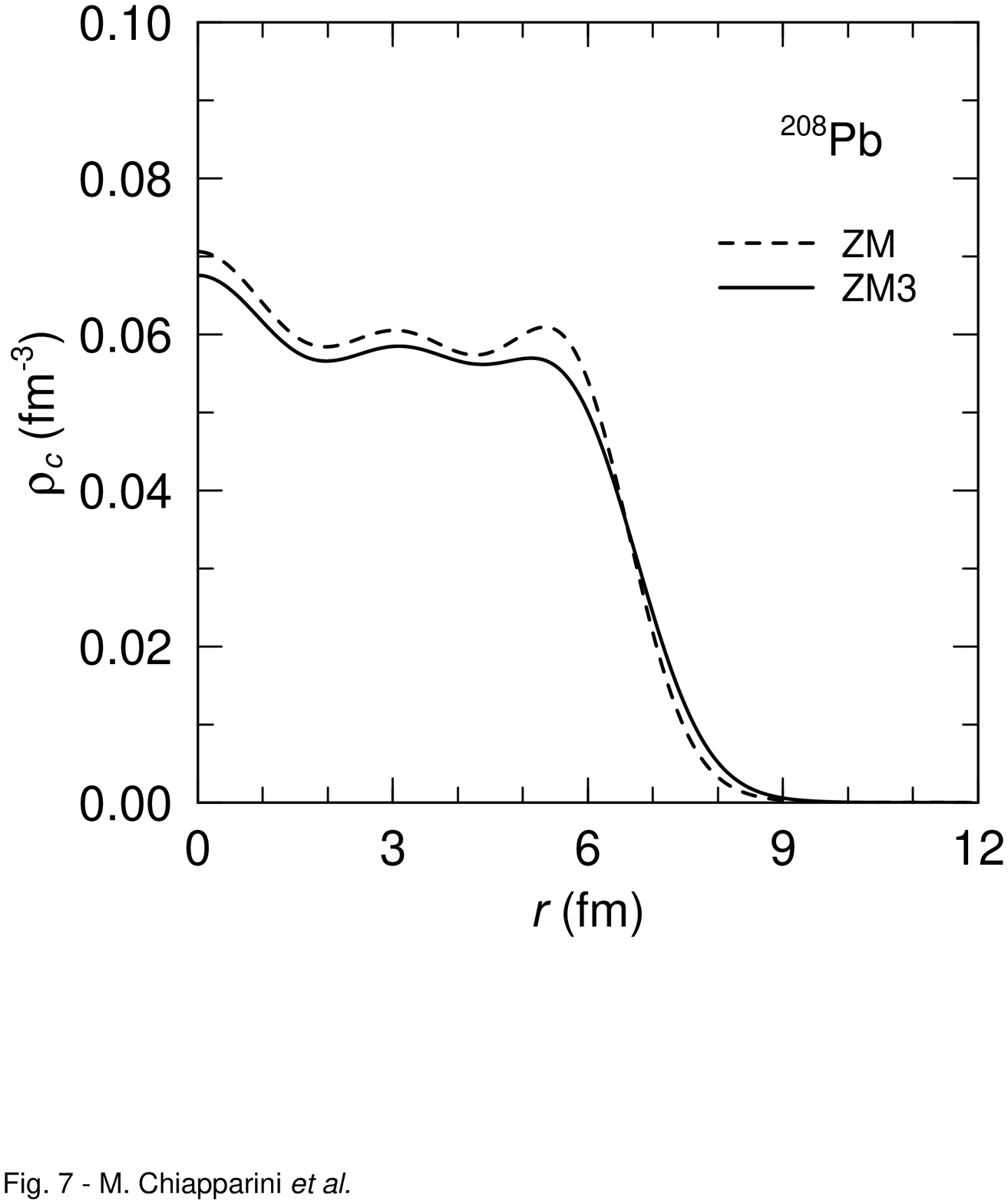}
\end{centering}

\newpage

\begin{centering}
\epsfysize=21.5truecm
\epsffile{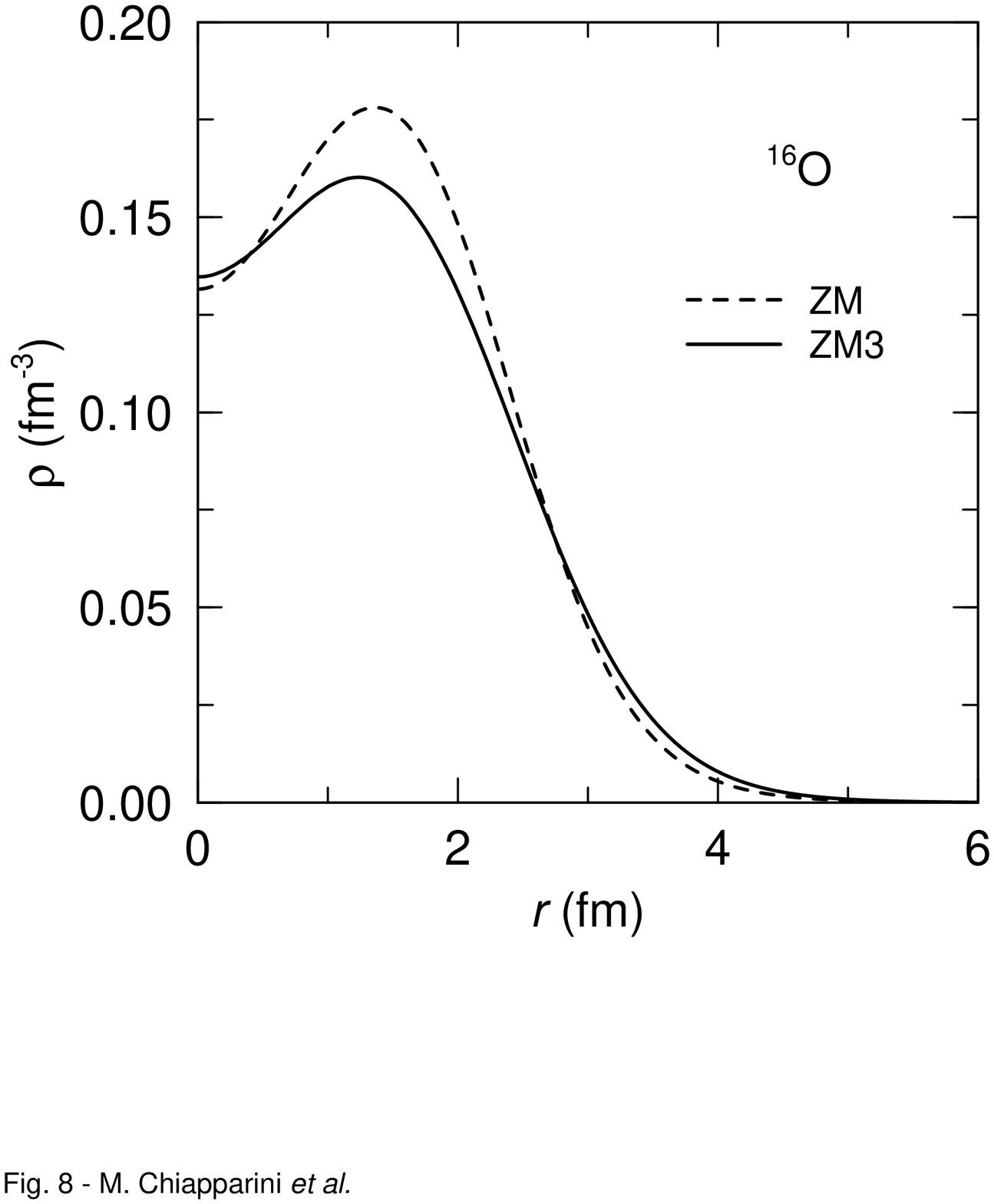}
\end{centering}

\newpage

\begin{centering}
\epsfysize=21.5truecm
\epsffile{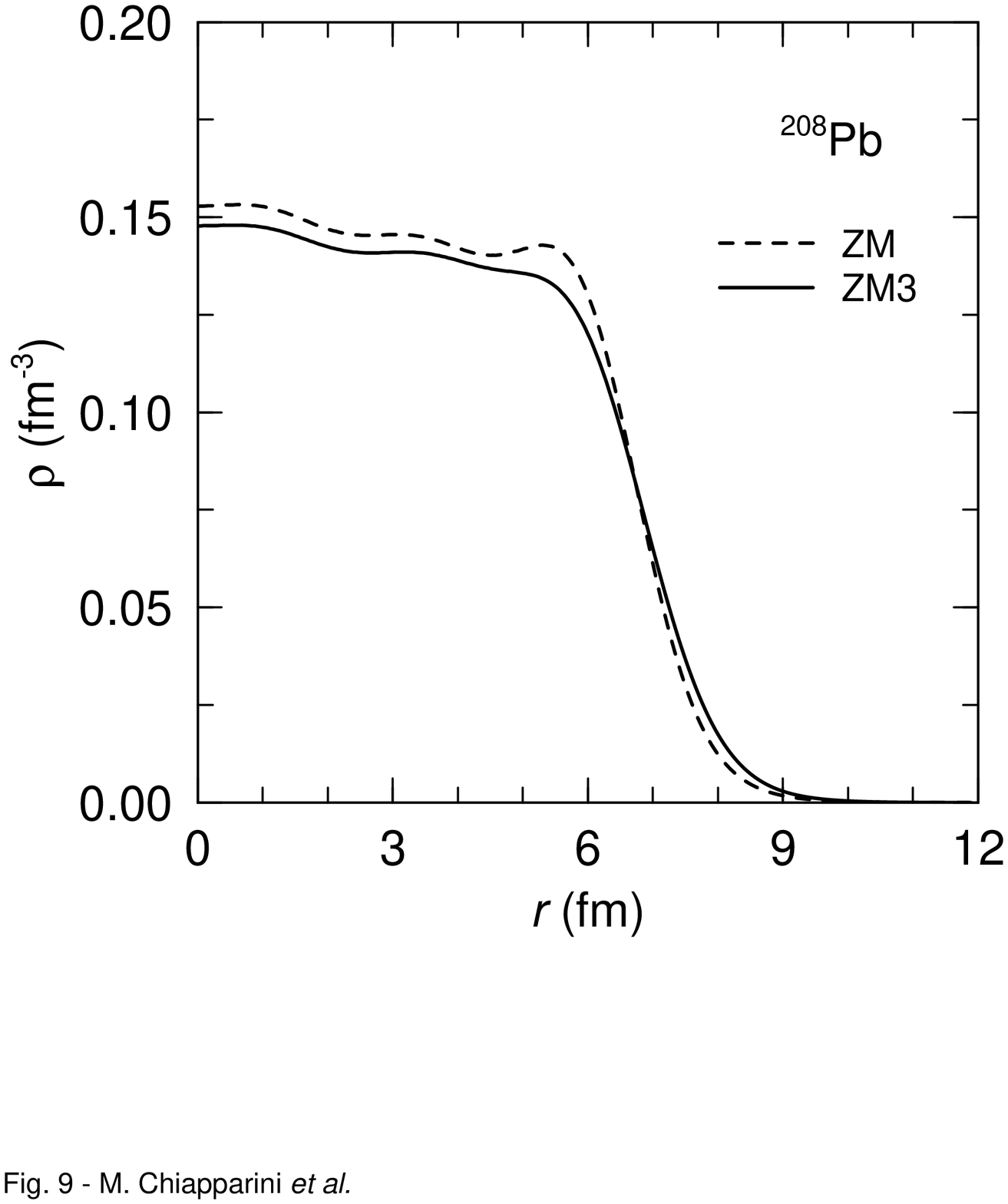}
\end{centering}

\end{document}